\documentclass[final,times,twocolumn]{elsarticle}
\pdfoutput=1
\journal{Computer Networks}

\usepackage{amssymb}

\usepackage{amsmath}
\usepackage{algorithmicx}

\usepackage{lipsum}

\usepackage{times}
\usepackage{url}
\usepackage{multirow}
\usepackage{algpseudocode}
\usepackage{algorithm}
\usepackage{rotating}
\usepackage{subcaption} 
\usepackage{commath}
\allowdisplaybreaks

\usepackage[table]{xcolor}
\usepackage{color, colortbl}
\definecolor{Gray}{gray}{0.9}

\newif\ifcoloractive

\definecolor{LightCyan}{rgb}{0.88,1,1}
\algnewcommand\algorithmicinput{\textbf{INPUT:}}
\algnewcommand\INPUT{\item[\algorithmicinput]}
\algnewcommand\algorithmicoutput{\textbf{OUTPUT:}}
\algnewcommand\OUTPUT{\item[\algorithmicoutput]}
\algdef{SE}[DOWHILE]{Do}{doWhile}{\algorithmicdo}[1]{\algorithmicwhile\ #1}

\makeatletter
\newcounter{algorithmbis}
\renewcommand{\thealgorithmbis}{\arabic{algorithmbis}}
\def\algorithmbis{\@ifnextchar[{\@algorithmbisa}{\@algorithmbisb}}
\def\@algorithmbisa[#1]{%
  \refstepcounter{algorithmbis}
  \trivlist
  \leftmargin\z@
  \itemindent\z@
  \labelsep\z@
  \item[\parbox{0.49\textwidth}{%
    \hrule
    \noindent\strut\textbf{Algorithm \thealgorithmbis} #1
    \hrule
  }]\hfil\vskip+0em%
}
\def\@algorithmbisb{\@algorithmbisa[]}

\makeatother

\begin{document}
	\begin{frontmatter}

\title{Joint Reliability-aware and Cost Efficient Path Allocation and VNF Placement using Sharing Scheme}

\author[label2]{Abolfazl Ghazizadeh}
\address[label1]{School of Electronic Engineering and Computer Science, Queen Mary University of London, London, UK}
\address[label2]{Department of Electrical and Computer Engineering, Tarbiat Modares University, Tehran, Iran}
\ead{abolfazlghazizadeh@modares.ac.ir}
\author[label2]{Behzad Akbari}
\ead{b.akbari@modares.ac.ir}
\author[label1]{Mohammad M. Tajiki}
\ead{m.tajiki@qmul.ac.uk}

\begin{abstract}
{Network Function Virtualization (NFV) is a vital player of modern networks providing different types of services such as traffic optimization, content filtering, and load balancing.} More precisely, NFV is a provisioning technology aims at reducing the large Capital Expenditure (CAPEX) of network providers by moving services from dedicated hardware to commodity servers using Virtualized Network Functions (VNF). A sequence of VNFs/services following a logical goal is referred to as a \textit{Service Function Chain (SFC)}. The movement toward SFC introduces new challenges to those network services which require high reliability. To address this challenge, redundancy schemes are introduced. Existing redundancy schemes using dedicated protection enhance the reliability of services, however, they do not consider the cost of redundant VNFs. In this paper, we propose a novel reliability enhancement method using a shared protection scheme to reduce the cost of redundant VNFs. 
To this end, We mathematically formulate the problem as a Mixed Integer Linear Programming (MILP). The objective is to determine optimal reliability that could be achieved with minimum cost. Although the corresponding optimization problem can be solved using existing MILP solvers, the computational complexity is not rational for realistic scenarios. Thereafter, we propose a Reliability-aware and minimum-Cost based Genetic (RCG) algorithm to solve this problem with low computational complexity. In order to evaluate the proposed solution, We have compared it with four different solutions. Simulation results show that RCG achieves near-optimal performance at a much lower complexity compared with the optimal solution.
\end{abstract}

\end{frontmatter}
\begin{keyword}
Software Defined Network (SDN); Network Function Virtualization (NFV); Service Function Chaining (SFC); Fault Tolerance; Redundancy Scheme; Resource Reallocation.
\end{keyword}

\section{Introduction}\label{sec:introduction}
    Network traffic flows may need to be served or screened through different hardware middle-boxes while passing the network; as an example of such middle-boxes consider HTTP proxies, Intrusion Detection Systems (IDSs), Network Address Translators (NATs), and firewalls. In order to reduce the capital and operational expenditure of using middle-boxes and to increase the flexibility and scalability of services provided by them, Network Function Virtualization (NFV) replaces hardware middle-boxes with more flexible software applications known as \textit{Virtual Network Functions (VNFs).} On the other hand, the Software Defined Networking (SDN) paradigm offers the possibility to control the forwarding of packets from a logically centralized point of view, thus easing the introduction of efficient and flexible algorithms to optimize the utilization of network and processing resources~\cite{zhang2016co}. Motivated by the collaboration of SDN and NFV, the topic of VNF as a Service (VNFaaS) is currently under attentive study by both telecommunication and cloud stakeholders as a promising direction~\cite{casazza2019availability}.
    
    Optimal resource allocation is an essential metric for network providers to reduce their costs and maximize their efficiency~\cite{tajiki2016qrtp}. Besides, to increase customers' Quality of Experience (QoE) and minimize the energy consumption, the VNFs need to be be dynamically relocated between network nodes, i.e., a running VNFs may need to migrate from a server to another one. Consequently, the placement of VNFs is a fundamental issue to efficiently deploy NFV technology. On the other hand, to optimize the resource utilization on both nodes and links, developing efficient algorithms for the joint problem of VNF placement and network traffic routing becomes an essential step~\cite{nguyen2019optimizing}.
    
    Another important metric of choosing a service provider is the reliability of its services. This forces the service providers to seek for NFV deployment algorithms that keep the reliability above some standards. VNFs are usually executed on commercial-off-the-shelf (COTS) network elements. COTS elements are characterized as low reliable devices meaning their reliability is significantly lower than carrier-grade equipments. Additionally, the COTS's operation may be affected by increasing the computing load, hardware failures  or malicious attacks~\cite{kang2017trade}. To ensure a desired level of end-to-end (e2e) reliability, redundancy scheme is an efficient way that is used in many works. There are two types of redundancy: 1: with dedicated protection, 2: with shared protecting. Existing redundancy methods with dedicated protection, enhance the reliability of services without considering the cost of redundant network functions. On the other hand, existing redundancy methods with shared protecting use an On-demand scheme that increases preparation time up to 3 times~\cite{cziva2017container}. 
    
    Motivated by the aforementioned considerations, we address the joint problem of VNF placement and flow routing with reliability and QoS considerations. More precisely, we study the joint problem with the objective of maximizing the resource utilization while keeping the reliability in a desirable threshold using a minimum set of redundant functions. We only consider the reliability of the computational node, because link reliability issues can easily be converted to node reliability. To this end, we exploit redundancy schemes by mathematically formulating the the problem of minimum resource consumption with respect to QoS constraints. Thereafter, we use an Mixed Integer Linear Programming (MILP) solver to optimally solve the corresponding optimization problem. Due to the high computational complexity of MILP solvers, we propose an efficient meta-heuristic algorithm to handle the scalability issue over large-scale networks. Our main contributions are summarized as follows:
    \begin{itemize}
        \item We propose a new reliability-aware resource allocation algorithm using shared protection scheme with Active-Standby redundancy. The algorithm is proposed for software defined networks to address the SFC problem with the objective of minimizing redundant VNFs without affecting the Quality of Service (QoS) parameters;
        \item Mathematical formulation of the joint problem of VNF placement and routing for the proposed protection scheme by considering QoS parameters. The corresponding optimization problem belongs to the class of mixed-integer quadratically constrained programming (MIQCP) in our first natural formulation;
        \item Linearization of the non-linear constraints in order to have the modeling in form of Mixed integer linear programming (MILP) which is solvable using existing ILP solvers such as IBM CPLEX;
        \item We propose a near optimal meta-heuristic algorithm to solve the mentioned problem in a reasonable execution time. The proposed algorithm is an scalable solution which can be used for large-scale networks;
        \item Comparison of the Genetic algorithm with state-of-the-art algorithms and the optimal solution through a set of various metrics, which includes: i) execution time, ii) bandwidth consumption, and iii) transmission latency.
    \end{itemize}
  
    The rest of this paper is organized as follows: Section~\ref{sec:relatedwork} goes through literature and surveys related works. Section~\ref{sec:sharedprotection} discusses one of the most important reliability enhancement schemes called 'shared protection scheme' and compares it with the other schemes using illustrative examples. Section~\ref{sec:formulation} then provides the system model and problem formulation. To solve the scalability issues a meta-heuristic algorithm is proposed which is described in Section~\ref{sec:GA}. Besides, to evaluate the proposed solution, numerical results are presented in Section~\ref{sec:results}. Finally, the paper is concluded and Remarks and outlines regarding the open research problems are included in section~\ref{sec:conclusions}.
    
\section{Related Works}\label{sec:relatedwork}
In the following, the main literature on NFV related to our work is discussed. From now on we refer to the Joint problem of path Allocation and VNF placement as \textit{Service Function Chaining (SFC)}. Related works are divided into three different categories: i) SFC solutions unaware of reliability \cite{bhamare2017optimal,bari2015orchestrating,even2016approximation,tajiki2018joint,xu2015effective,rocha2015network,zhao2015joint,calcavecchia2012vm,meng2010improving,eramo2017migration}, ii) SFC solutions focusing on minimizing the fault/failure probability \cite{tajiki2019joint, vilchez2014self,fonseca2017survey,sterbenz2010resilience,tajiki2019software,kreutz2013towards,sharma2013openflow, van2014fast,tajiki2018energy}, and iii) SFC solutions focusing on redundancy protection \cite{virtualisation3reliability,fan2015grep, ye2016joint, carpio2017vnf,pham2017online,kang2017trade,qu2017reliability}. We then describe the works falling in each category. 
\subsection{SFC solutions unaware of reliability}\label{sec:relatedwork.1}
There are numerous works focusing on providing SFC in software defined networks. An SFC taxonomy that considers performance and architecture dimensions is introduced in \cite{medhat2017service} which could be used as the basis for the subsequent state-of-the-art analysis.

The authors of \cite{bhamare2017optimal} exploit NFV architecture to deploy SFC. Specifically, in their work, the problem of VNF placement for the optimal SFC formation in a multi-cloud scenario is investigated. Moreover, they set up the problem of minimizing inter-cloud response time and traffic across geographically distributed data centres as an ILP optimization problem, along with some other constraints such as total SLA and deployment costs.
Moreover, in \cite{reddy2016robust} an optimization model based on the concept of $\Gamma$-robustness is proposed. They focus on dealing with the uncertainty of the traffic demand. The authors of \cite{zhang2016co} propose a heuristic algorithm to find a solution for service chaining. It employs two-step flow selection when an SFC with multiple network functions needs to scale out. Furthermore, the authors in \cite{abdelsalam2017implementation} introduce a VNF chaining which is implemented through segment routing in a Linux-based infrastructure. To this end, they exploit an IPv6 Segment Routing (SRv6) network programming model to support SFC in an NFV scenario. The authors of \cite{kulkarni2017neo} propose a scheme which provides flexibility, ease of configuration and adaptability to relocate the service functions with a minimal control plane overhead.
    
Besides, the authors of \cite{bari2015orchestrating} use ILP to determine the required number and placement of VNFs that optimize network CAPEX/OPEX costs without violating SLAs. In \cite{even2016approximation} an approximation algorithm for path computation and function placement in SDNs is proposed. Similar to \cite{bari2015orchestrating}, they proposed a randomized approximation algorithm for path computation and function placement. In \cite{ghaznavi2016service} an optimization model to deploy a chain in a distributed manner is developed. Their proposed model abstracts heterogeneity of VNF instances and allows them to deploy a chain with custom throughput without worrying about individual VNFs’ throughput. The paper \cite{rost2016service} considers the offline batch embedding of multiple service chains. They consider the objectives of maximizing the profit by embedding an optimal subset of requests or minimizing the costs when all requests need to be embedded.

\subsection{SFC solutions focusing on minimizing the fault/failure probability} \label{sec:relatedwork.2}

The available literature ranges from the problem of fault detection and recovery solutions ~\cite{vilchez2014self,fonseca2017survey} to the problem of fault-aware routing of the network traffic in SDN/NFV infrastructure \cite{sterbenz2010resilience,kreutz2013towards}. 
In detail, in \cite{kreutz2013towards} the authors focus on analyzing the fault tolerance in SDN. More in detail, they discuss failure occurrence and fault tolerance in the OpenFlow-enabled networks. The main goal is to propose a node/link failure recovery and fault detection method in the data plane that can be controlled through the controller. However, they neither cover the SFC fault-awareness, nor consider the application plane side-effect. 

In \cite{sharma2013openflow}, controller-based fault recovery solution focusing on pre-configured backup paths and path-failure detection is presented. More precisely, they present a solution to manage the traffic flows by pre-configuring the network, which is not an effective solution, by design, in real scenarios. The authors of \cite{van2014fast} propose a cost-efficient solution to detect link failures in order to increase the fault tolerance by combining the flow retrieval which is achieved through analyzing the protection switching times and using a fast protection method. Interestingly, this paper supports the fault minimization over the links and addresses the end-to-end fault tolerance method per flow, but the solution is not secured against occurrence of failure. In fact, the system tries to minimize the probability of failure but it cannot handle the occurrence of failure. 
The authors in \cite{tajiki2019joint}, present an architecture for Fault Prevention and Failure Recovery which is a multi-tier structure in which the network traffic flows pass through networking nodes to decrease the energy consumption and network side-effects of traffic engineering. Similar approach is taken in \cite{tajiki2019software}, to formulate the problems of flow routing, allocation of VNFs to flows, and VNF placement as Integer Linear Programming optimization problems. Since the formulated problems cannot be solved in acceptable timescales for real-world problems, they propose several cost-efficient and quick heuristic solutions. Both \cite{tajiki2019software} and  \cite{tajiki2019joint} reduce the probability of failure in physical servers, however, they both expose the network unprotected in case of failure in a networking node.

\subsection{SFC solutions focusing on redundancy protection} \label{sec:relatedwork3}
Numerous works focus on  increasing the reliability of each service/VNF separately and do not take the advantages of considering the global information of the VNF Forwarding Graph (VNF-FG). The main drawback of focusing on services/VNFs separately is low utilization of networking resources. A survey of the recent works on SFC is presented in~ \cite{virtualisation3reliability} classifying VNF/service protection into three groups: Active-Standby, Active-Active, and on-demand. In the following, some of the state-of-the-art solutions proposed for redundancy protection are discussed briefly.
In \cite{fan2015grep}, an algorithm for minimizing the physical resources consumption is proposed which guarantees the required reliability with polynomial time complexity. The proposed scheme ignores the global information of the VNF-FG and cost of backups, which leads to the VNF over-replication.

An on-demand scheme is a lazy approach of tackling the VNFs failure meaning that it postpones the resource allocation of the backup function to a later time when the failure has occurred.
In \cite{fan2015grep} and \cite{ye2016joint} authors used this method for enhancing reliability of services. This is an efficient way to improve the performance of resources, but increases the fault recovery time. 
In Active-Active scheme, all node (including redundant nodes) are active are serving incoming requests \\cite{virtualisation3reliability}. This solution not only requires redirecting traffic in case of failure but also requires a load balancer to be deployed in front of several backups.
The authors of~\cite{carpio2017vnf}, study the the potential of VNFs replications to accelerate network load balancing. In this way, they consider the problem of VNF placement with replications. They mathematically formulate their problem and propose three solution for the allocation and replication of services/VNFs: Genetic Algorithm (GA), LP solver, and Random Fit Placement Algorithm (RFPA). Similarly, in~\cite{pham2017online}, the optimization problem of load balancing is formulated as a mixed integer linear program. Thereafter, in order to solve the online load balancing problem a fast algorithm is developed.
In both~\cite{carpio2017vnf} and~\cite{pham2017online} authors focus on increasing the reliability using a replication data flow method through migrating backup functions from low reliability nodes to more reliable nodes. In this method while recovery time is very low the performance is not comparable with other existing methods.

Active-Standby is a method where active VNFs provide specific services, and these active VNFs are protected by one or more standby VNF(s). These redundant VNFs do not actively provide service and they require a mechanism to redirect traffic to them in case of failure. As an example, authors of \cite{kang2017trade} follows the Active-Standby method by seeking for a trade-off between end-to-end reliability and computational load over servers. In this way, they exploit the joint design of VNF Chain Composition (CC) and Forwarding Graph embedding (FGE) using a dedicated redundancy scheme. They model the problem in the form of Mixed-Integer LP (MLIP) and exploit existing toolbars to solve the problem. In the same way, authors of \cite{qu2017reliability} propose a multi-path backup scheme to enhance reliability while minimizing the end-to-end delay. Although the aforementioned schemes have many benefits, they lead to wasted resources since they focus on increasing the reliability of each service/VNF individually instead of considering the whole nodes as an integrated entity.

\section{The Propose Shared Protection scheme}\label{sec:sharedprotection}
Considering the fact that hardware components fail frequently due to various human and natural causes (earthquakes, malicious attacks, fibre cuts, etc.), network operators must use protection methods to provide reliable services/functions~\cite{han2015network}. Dedicated Protection (DP) scheme is a traditional way to enhance the reliability of SFCs. In this scheme, one or more redundant VNFs will be kept reserved for a service/function that needs high reliability. If DP is provisioned, the reliability of given VNF  can be obtained as follow:
    \begin{align}
        &r_i=1-(1-r^p_i)\cdot (1-r_b)\label{eqR1}
    \end{align}
where $r_i^p$ and $r_b$ are corresponding to the reliability of primary VNF and reliability of the backup VNF. 
Although DP can provide high reliability for services, it suffers from high usage of bandwidth and computational resources. In order to balance between resource utilization and reliability the Shared Protection (SP) is a well-known scheme. In this scheme, each backup function can be reserved for several primary functions. Using SP, the reliability of VNF $i$ is:
\begin{align}
    &r_i=r_{i}^p+(1-r_i^p)\cdot r_b\cdot\varphi_i\label{eqR2}\\
    &\varphi_i=1-\sum_{i\neq j}{\frac{MTTR_j}{MTTR_i+MTTR_j}\cdot (1-r_j)}\nonumber
\end{align}
where $r_i^p$ and $r_b$ are the initial reliability of primary and backup VNFs, respectively. $\varphi_i$ is the probability that the shared backup VNF can be assigned to VNF $i$~\cite{zhang2010availability}.

\begin{figure*}[!htbp]
    \centering
        \begin{subfigure}{0.33\textwidth}
    	    \centering
    		\includegraphics[width=1\linewidth]{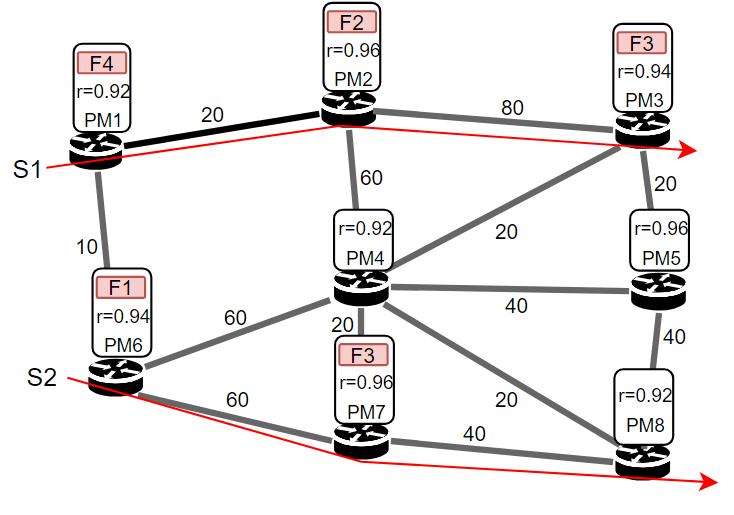}
        	\caption{No protection.}
        	\label{fig:protscha}
    	\end{subfigure}
    	\begin{subfigure}{0.33\textwidth}
    	    \centering
    		\includegraphics[width=1\linewidth]{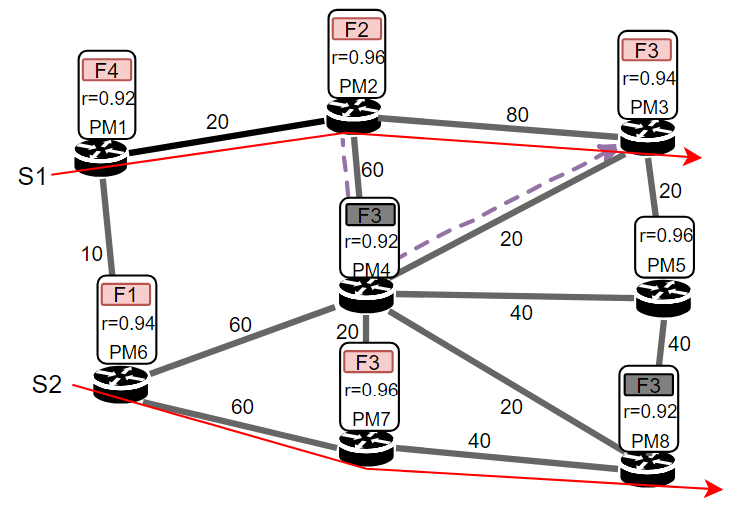}
        	\caption{Dedicate protection.}
        	\label{fig:protschb}
    	\end{subfigure}
    	\begin{subfigure}{0.33\textwidth}
    	    \centering
    		\includegraphics[width=1\linewidth]{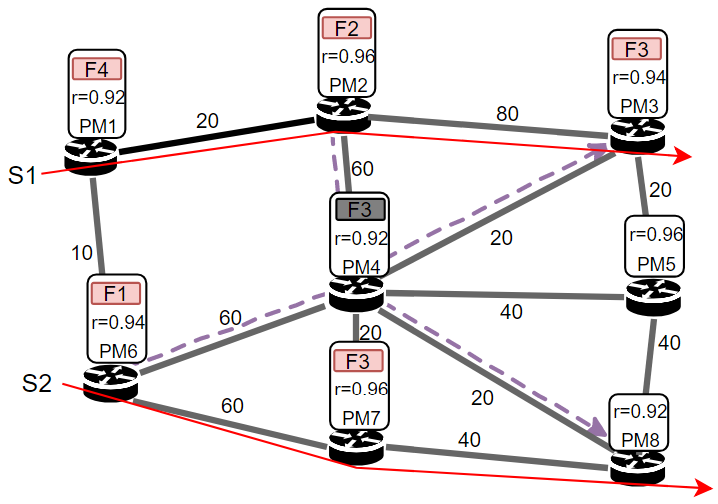}
        	\caption{Shared protection.}
        	\label{fig:protschc}
    	\end{subfigure}
    	\caption{Protection methods.}
        \label{fig:protsch}
    \end{figure*}

In order to clarify this method, we have given an example that compare reliability and bandwidth consumption in SP and DP. Also to evaluate the performance of proposed scheme, we mentioned a No Protection (NP) scheme. Consider a sub-network consists of eight Physical Machine (PM), namely $PM_1$ through $PM_8$ as illustrated in Fig.~\ref{fig:protscha}. The substrate network is assumed to host two service function chains, namely $s_1$ and $s_2$. $s_1$ requests for three functions consists of $\{f_4, f_2, f_3\}$ which are respectively hosted on $\{PM_1, PM_2, PM_3\}$ (initiated at $PM_1$ and destined to $PM_3$). Similarly, $S_2$ requests for two functions consist of $\{f_1, f_3\}$ which are respectively hosted on $\{PM_6, PM_7\}$ (initiated at $PM_6$ and destined to $PM_8$). Bandwidth requirements for each service is considered to be 20 units. Fig.~\ref{fig:protscha} illustrates No Protection scheme where the reliability of $s_1$ and $s_2$ are $r_{s_1}=0.94\times 0.96\times 0.92 = 0.83$ and $r_{s_2}=0.96\times 0.92=0.883$, respectively, and the consumed bandwidth is $b=80$ unit. Another example is shown in Fig.~\ref{fig:protschb}, where the $f_3$ of $s_1$ and $f_3$ of $s_2$ replicated into $PM_4$ and $PM_8$, respectively. In case of using DP, the reliability of $s_1$ and $s_2$ are $r_{s_1}=0.94\times 0.96\times(1-(1-0.94)\times (1-0.92)) = 0.898$ and $r_{s_2}=0.96\times(1-(1-0.92)\times(1-0.94))=0.955$, respectively, as well as the consumed bandwidth is $b=120$ unit.

Fig.~\ref{fig:protschb} illustrates the same scenario when DP scheme is deployed to ensure high reliability. The main disadvantage of SP is that, despite higher reliability obtained, it considerably increases the amount of the required resources. In order to reduce the number of replicated VNFs while holding the level of reliability, we purpose a shared protection scheme with Active-Standby redundancy. An example of SP illustrated in Fig.~\ref{fig:protschc} where one backup VNF type $f_3$ is placed on $PM_4$ and reserved for $f_3$ of $s_1$ and $f_3$ of $s_2$, simultaneously. According to Eq.~\ref{eqR2} the achieved services reliability in this case is:
\begin{align}
    &r_{s_1}=0.94\times 0.96\times 0.992=0.895\nonumber\\
    &r_{s_2}=0.94\times 0.992=0.932\nonumber
\end{align}
The consumed bandwidth is $b=160$ unit which is increased by 20$\%$ compared to DP.
   
\section{Problem Formulation}\label{sec:formulation}
    In this section, the SFC-aware resource allocation with respect to system reliability is presented. The system model is for a joint problem of VNF placement and flow routing. Consequently, it guarantees the best possible end-to-end reliability for the assigned path to each flow. We also consider the cost of using redundant resources by making a trade-off between reliability and cost.  We have QoS as a fundamental metric in our system model. Hence, the propose system not only ensures the required service of each flow to be delivered via the selected path but also the QoS of the service to be kept in a proper range. In the following, we detail the formulation used in the proposed formulation. Table~\ref{tab:notation} defines the symbols with a brief description.
    
    \begin{table*}[t]
        \caption{Main Notation.}\label{tab:notation}
        \centering
        \scriptsize
        \rowcolors{2}{gray!25}{white}
        \begin{tabular}{|c|l|}    
        \hline
            \textbf{Symbol} & \textbf{Definition}  \\\hline\hline
            \multicolumn{2}{|c|}{\textbf{Input Parameters}}\\\hline
            $\mathcal{N}$ & The set of Servers \\\hline
            $F_{i}$ & An ordered chain of VNFs corresponding to service chain $i$ \\\hline
            $\mathcal{F\prime}$ & The set of backup VNFs \\\hline
            $S$ & The set of network services \\\hline
            $L$ & The set of physical links \\\hline
            $T$ & The set of VNF’s Types \\\hline
            $c_{i,j}$ & The processing capacity requirement of $j$-th function of service $i$ \\\hline
            $\theta^i_{req}$ & The Minimum reliability accepted by service $i$ \\\hline
            $\phi^i_{req}$ & The Maximum delay accepted by service $i$ \\\hline
            $\phi_{l}$ & The delay of physical link $l$ \\\hline
            $\theta_{k}$ & The reliability of server $k$\\\hline
            $\sigma_{i}$ & The source of network service $i$\\\hline
            $\delta_{i}$ & The Destination of network service $i$\\\hline
            $B_{m}$ & The bandwidth capacity of link $m$ \\\hline
            $C_{k}$ & The processing capacity of server $k$ \\\hline
            $x_{i,j}$ & The Type of $j$-th function of service $i$\\\hline
            $x^\prime_{l}$ & The Type of backup function $l$ \\\hline
            \multicolumn{2}{|c|}{\textbf{Variables}}\\\hline
            $y^k_{i,j}$ & A binary variable that equals 1 if and only if $j$-th function of service $i$ is located on server $k$ \\\hline
            $y^\prime k_{i}$ & A binary vari, \eqref{eqR12}ble that equals 1 if and only if $i$-th backup is located on server $k$\\\hline
            $U^l_{i, j}$ & A binary variable that equals 1 if and only if $l$-th backup is assigned to $j$-th function of service $i$ \\\hline
            $W^m_{i, j}$ & A binary variable that equals 1 if and only if service $i$ use link $m$ for access to $j$-th VNF \\\hline
            $W^\prime l_{i,j}$ & A binary variable that equals 1 if and only if service $i$ use link $j$ for access to backup $l$\\\hline
        \end{tabular}
    \end{table*}
    
    Consider a substrate network as a directed graph $G=(N,L)$, which consist of a set of physical machines $N$ and directed links $L$. Let $C_k$  be the processing capacity of $PM_k$ where $k\in N$ and each $PM$ can execute several VNFs, depend on its $C_k$. Let $B_m$ donate the bandwidth capacity of link $m$ where $m\in L$. We donate by $S$ a set of demanded services. Each service $s_i\in S$ is specified by a required bandwidth $b_i$ and source node  $\sigma_i$ and destination node $\delta_i$. Let $F_i$ be the ordered chain of VNFs corresponding to service chain $s_i$.

    In the following, we develop a Mixed Integer Linear Programming (MILP) model to mathematically formulate the problem of reliability enhancement with shared protection scheme. We present the MILP model with all the notation specified in Table~\ref{tab:notation}.In order to make the understanding of mathematical formulation easier, the model is divided into seven parts and each part is discussed separately.

\subsection{Reliability constraints}\label{subsec:reliability} 
    In this part, constraints related to reliability are discussed. The operation of each VNF may be affected by unexpected failure in its software or its physical machine ($PM$). Each $PM$ has specific values for its Mean Time Between Failure (MTBF) and Mean Time To Repair (MTTR). 
    For the sake of simplicity, we refer to the j'th VNF of service i as $VNF_{i,j}$. Let $r^{p}_{i,j}$ be the reliability of one instance of $VNF_{i,j}$. This value is highly influenced by the reliability of the $PM$ hosting this VNF (Eq.~\ref{eqR3}). Similarly, $r_{i,j}$ is the reliability of $VNF_{i,j}$ considering all instances including the backups.
	\begin{align} 
	    &r_{i,j}=r^{p}_{i,j}+\left(1-r^{p}_{i,j}\right)\times r_l\times U_{i,j}^l\times\nonumber\\ 
	    &\left[1-\sum_{i\neq i^\prime}{\sum_{j^\prime\in \left[1,\abs{F^\prime_{i^\prime}}\right]}{\frac{MTTR_{i^\prime j^\prime}}{MTTR_{i,j}+MTTR_{i^\prime j^\prime}}\times \left(1-r_{i^\prime j^\prime}\right)\times U_{i^\prime j^\prime}^l}}\right]\nonumber\\ 
	    &\forall i,i^\prime\in \left[1,\abs{S}\right],j,j^\prime\in \left[1,\abs{F_i}\right],l\in \left[1,\abs{F^\prime}\right]\label{eqR3}
    \end{align}

    Let $\theta^i_{req}$ be the minimum reliability accepted by service i. according to \ref{eqR3}, the achieved reliability $\theta^i$ of an arbitrary network service $s_i$ is given by:

    \begin{align}
        &\theta^i = \prod_{j\in\left[1,\abs{F_i}\right]}{r_{i,j}}, ~\forall i\in \left[1,\abs{S}\right],j\in \left[1,\abs{F_i}\right]\label{eqR4}
    \end{align}
    
    Where $r_{i,j}$ is the reliability of $j^{th}$ VNF of service $s_i$. If achieved reliability $\theta^i<\theta_{req}^i$, then improve reliability of services $i$ with add one or more backups to its primary VNFs. For all of network services, the number of redundant VNFs should be sufficient to satisfy its reliability requirement.
    Let $F^\prime$ denote the set of backup VNFs. Each redundant VNF may be shared with several primary VNF. If redundant VNF $l$ is assigned to VNF j of service $s_i$, $U_{i,j}^l$ will be 1 otherwise 0. Let $x_l^\prime$ and $x_{i,j}$ respectively be the type of redundant VNF $l$ and the type of $j^{th}$ VNF of service $s_i$. The below constrain allow each primary VNF to use redundant VNFs which is the same type.

    \begin{align}
        &x^\prime_l\cdot x_{i,j} \geq U_{i,j}^l, \forall i\in \left[1,\abs{S}\right], j\in \left[1,\abs{F_i}\right], l\in \left[1, \abs{F^\prime}\right] \label{eqR5}
    \end{align}
    
    \subsection{Routing Constraints}
    In the following, the constraints for flow routing with respect to QoS are discussed. In this formulation, m.head and m.tail respectively represent the first and the last node along link m.
    \begin{align}
        &\sum_{m\in L \And  m.tail=\sigma_i}{W_{i,j}^m=1} \label{eqR6}\\
        &\sum_{m\in L \And m.head=\delta_i}{W_{i,j}^m=1}
        \label{eqR7}\\
    	&\forall i\in \left[1,\abs{S}\right],j\in \left[1,\abs{F_i}\right]\nonumber
    \end{align}
    where Eq.~\ref{eqR6} and~\ref{eqR7} make sure that the path of each service starts from $\sigma_i$ and ends in $\delta_i$, precisely.
    
    \begin{align}
        &\sum_{m\in L \And m.head=k}{W_{i,j}^m\times b_{i,j}} = \sum_{n\in L \And n.tail=k}{W_{i,j}^n\times b_{i,j}}\label{eqR8}\\
        &\forall  i\in \left[1,\abs{S}\right],j\in \left[1,\abs{F_i}\right], k\in \left[1,\abs{N}\right]\nonumber
    \end{align}
    Eq.~\ref{eqR8} ensures that for each service, the amount of input load to each server is equal to the amount of its output load. Unless the server is the first node (start node) or the last node (end node) of that service.
    
    \begin{align}
        &\sum_{m\in L \And m.tail=k}{{W^\prime}_{i,m}^l} \geq U_{i,j}^l\times{y^\prime}_l^k \label{eqR9}\\
    	&\sum_{m\in L \And m.head=k}{{W^\prime}_{i,m}^l} \geq U_{i,j}^l\times {y^\prime}_l^k\label{eqR10}\\
    	&\forall i\in \left[1,\abs{S}\right],j\in \left[1,\abs{F_i}\right],k\in \left[1,\abs{N}\right],l\in \left[1,\abs{F^\prime}\right]\nonumber
    \end{align}
    Eq.~\ref{eqR9} and~\ref{eqR10} make sure that if backup VNF $l$ is assigned to one of the functions of the service $i$, there is a backup path that passes through the server which hosts the VNF~$l$.
    
    \begin{align}
        &\sum_{m\in L \And m.tail=k}{{W^\prime}_{i,m}^l} \geq U_{i,j}^l\times y_{i,j-1}^k\label{eqR11}\\
    	&\sum_{m\in L \And m.head=k}{{W^\prime}_{i,m}^l} \geq U_{i,j}^l\times y_{i,j+1}^k\label{eqR12}\\
    	&\sum_{m\in L \And m.tail=k}{{W^\prime}_{i,m}^l} + y_{i,j+1}^k\geq  1\label{eqR13}\\
    	&\sum_{m\in L \And m.head=k}{{W^\prime}_{i,m}^l} + y_{i,j-1}^k\geq 1\label{eqR14}\\
    	& \forall i\in \left[1,\abs{S}\right],j\in \left[1,\abs{F_i}\right],k\in \left[1,\abs{N}\right],l\in \left[1,\abs{F^\prime}\right]\nonumber
    \end{align}
    In Eq.~\ref{eqR11}-\ref{eqR14} a path is marked as used to reach the backup $l$ by $j$'th VNF of service $i$ if the path Precisely starts from $(j-1)$ and ends in $(j+1)$.
    
    \begin{align}
        &\sum_{m,m^\prime\in L \And m.head=m^\prime.tail = k \And m^\prime.head = m.tail = k^\prime}{{W^\prime}_{i,m}^l + {W^\prime}_{{i,m}^\prime}^l} \leq 1\label{eqR15}\\
        &\forall i\in \left[1,\abs{S}\right],j\in \left[1,\abs{F_i}\right],k,k^\prime\in \left[1,\abs{N}\right],l\in \left[1,\abs{F^\prime}\right]\nonumber
    \end{align} 
    Eq.~\ref{eqR15} prevents the formation of the loops on the path and Eq.~\ref{eqR16} prevents the path from being cut off.
    
    \begin{align}
        &\sum_{m\in L \And m.tail=m^\prime.head}{{W^\prime}_{{i,m}^\prime}^l+y_{i,j-1}^{m.tail}} \geq {W^\prime}_{i,m}^l \label{eqR16}\\
       &\forall i\in \left[1,\abs{S}\right],j\in \left[1,\abs{F_i}\right],k\in \left[1,\abs{N}\right],l\in \left[1,\abs{F^\prime}\right]\nonumber
    \end{align}

\subsection{The NFV Placement and Anti-affinity constraints}\label{sec:NFV}
       	\begin{align}
       	    &\sum_{k\in \left[1,\abs{N}\right]}{y_{i,j}^k} =1,\ \ \forall i\in \left[1,\abs{S}\right],j\in \left[1,\abs{F_i}\right]\label{eqR17}\\
       	    &\sum_{k\in \left[1,\abs{N}\right]}{{y^\prime}_l^k} =1,\ \ \forall l\in \left[1,\abs{F^\prime}\right]\label{eqR18}
       	\end{align}
    Eq.~\ref{eqR17} and~\ref{eqR18} make sure that Each VNF, such as backup or primary, is executed by one and only one $PM$. As such, the Anti-affinity constraints are formulated as:
    \begin{align}
        &y_l^{\prime k}+y_{i,j}^k+U_{i,j}^l \leq 2\label{eqR19}\\ &\forall i\in \left[1,\abs{S}\right], j\in \left[1,\abs{F_i}\right],l\in \left[1,\abs{F^\prime}\right],k\in \left[1,\abs{N}\right]\nonumber
    \end{align}
    where Eq.~\ref{eqR19} ensures $U_{i,j}^l \neq 1$ if and only if both primary VNF and selected backup VNF are hosted by same $PM$. Because in the event of a Fail for a $PM$, only one of the primary function or backup function of a service fails.

    \begin{align}
        &y_{i,j}^k + y_{i^\prime,j^\prime}^k + U_{i,j}^l + U_{i^\prime,j^\prime}^l \leq 3\label{eqR20}\\
        \forall & i, i^\prime \in \left[1,\abs{S}\right], j\in \left[1,\abs{F_i }\right],j^\prime\in \left[1,\abs{F_i^\prime }\right],\nonumber\\&l\in \left[1,\abs{F^\prime}\right],k\in \left[1,\abs{N}\right], i\neq i^\prime\nonumber 
    \end{align}
    Eq.~\ref{eqR20} ensures that if VNF j and VNF $j^\prime$select one backup then the functions should be placed on different PMs. Because if they are located on the same PM, when the PM fails, they will need two backup functions at the same time.
\subsection{Bandwidth constraint} \label{subsec:band}
      We formulated the allocated bandwidth problem as:
      \begin{align}
          BW = &\sum_{i\in \left[1,\abs{S}\right]}\sum_{j \in \left[1, \abs{F_i}\right]}{\sum_{m\in \left[1,\abs{L}\right]}{W_{i,j}^m\times b_i}} +\label{eqR21}\\ &\sum_{l\in \left[1,\abs{F^\prime}\right]}{\sum_{m\in \left[1,\abs{L}\right]}{\max_{i\in \left[1,\abs{S}\right]}{\left(W_{i,m}^{\prime l}\times U_{i,j}^l\times b_i\right)}}}\nonumber
      \end{align}
      where $BW$ is the total allocated bandwidth and obtained from the sum of allocated bandwidth of each link which obtained from the sum of consuming bandwidth of services which selected the link and accumulate of maximum reserved bandwidth among the services that selected this link as a backup path to access the same backup VNF.
      
      \begin{align}
          &\sum_{i\in \left[1, \abs{S}\right]}{\sum_{j\in \left[1, \abs{F_i}\right]}{w_{i,j}^m\times b_i}} +\label{eqR22}\\
          &\sum_{l\in \left[1,\abs{F^\prime}\right]}{\max_{i\in \left[1,\abs{S}\right]}{\left(W_{i,m}^\prime l\times U_{i,j}^l\times b_i\right)}} < B_m,
          \forall m\in \left[1,\abs{L}\right]\nonumber
      \end{align}
      Eq.~\ref{eqR22} ensures that the total allocated bandwidth on any physical link $l$ cannot exceed its bandwidth capacity $B_m$.
\subsection{Computational capacity constraints} \label{subsec:comp}
    \begin{align}
        &\sum_{i\in \left[1,\abs{S}\right]}{\sum_{j\in \left[1,\abs{F_i}\right]}{{y_{i,j}^k\times C_{i,j} }}} +\label{eqR23}\\ &\sum_{l\in \left[1,\abs{F^\prime}\right]}{\max_{i\in \left[1,\abs{S}\right]}{\left(y_l^{\prime k}\times U_{i,j}^l\times C_{i,j}\right)}} < C_k, ~\forall k\in \left[1,\abs{N}\right]\nonumber
    \end{align}
    	Equation (23) ensures that the total allocated computing resources on any $PM_k$ cannot exceed its capacity $C_k$.
\subsection{Delay constraints}\label{subsec:dly}
    The delay constraints are formulated as follows:
    \begin{align}
        &\sum_{j^\prime \in \left[1,\abs{F_i}\right]\And j\neq j^\prime \And j\neq \left(j^\prime+1\right)}{\sum_{m\in \left[1,\abs{L}\right]}{W_{i,j^\prime}^m\times \phi_m}} +\label{eqR24}\\ &\sum_{m\in \left[1,\abs{L}\right]}{\sum_{l\in \left[1,\abs{F^\prime}\right]}{W^{\prime l}_{i,m}\times U_{i,j}^l\times \phi_m}} \leq {\phi^i}_{req}\nonumber\\
        &\forall i\in \left[1,\abs{S}\right], j\in \left[1,\abs{F_i}\right]\nonumber
    \end{align}
    where Eq.~\ref{eqR24} ensures the experienced delay for each service in any combination of primary path and backup path is less than the maximum delay accepted by the service.
\subsection{Objective Function}
    The objective function is establishing reliable service chains while minimizing the resource consumption. Our optimization problem is based on two objective:
    \begin{itemize}
        \item Minimizing the bandwidth usage caused by both primary and backup functions:
            $$\min{\left(\frac{BW}{\sum_{m\in [1,\abs{L}]}{B_m}}\right)}$$
            where $BW$ is the total allocated bandwidth and obtained from Eq.~\ref{eqR21} and $B_m$ is the bandwidth capacity of link $m$.
        \item Minimizing the utilization of the processing capacity by minimizing number of backup VNFs:
            $$\min{\left(\frac{\abs{F^\prime}}{\abs{F}}\right)}$$ 
            where $\abs{F^\prime}$ is number of backup functions and $\abs{F}$ is number of primary functions.
    \end{itemize}
    
    Our objective represents a Multi Criteria Decision Making. The comprehensive objective function is given by:
    $$\min{\left(\alpha\cdot \frac{\abs{F^\prime}}{\abs{F}}+(1-\alpha)\cdot\frac{BW}{\sum_{m\in [1,\abs{L}]}{}B_m}\right)}$$
    where $\alpha$ is the preference weight of each sub-goal. Coefficient $\alpha$ has a critical impact on the performance of the proposed solution. The selection of $\alpha$ is determined by two criteria: computational consumption and bandwidths consumption. A higher $\alpha$ implies that the  of VNFs computational consumption of the solution is closer to its optimal value, whereas a lower $\alpha$ implies that the bandwidths consumption is closer to its optimal value. Therefore, the resource utilization is parametric, this enables the datacenter owner to modify the minimization goal. For example, if the datacenter owner feels lack of available bandwidths, then a lower $\alpha$ can be assigned to the algorithm, which results in a less bandwidths consumption. If lack of computational resource is more sensitive, then a higher $\alpha$ can be assigned. This provides flexibility in respect to different perceptions about what needs to be more minimized. We empirically found that $\alpha=10$ jointly minimizes computational and bandwidths consumption across multiple datasets.
    The objective function considers the minimization of four different costs: reliability, server utilization, migration costs and link utilization.
  
\section{Genetic Algorithm}\label{sec:GA}
    Since using existing MILP solvers for the proposed formation is quite complex and challenging even for medium-scale networks, we propose a genetic algorithm to practically solve it.
    In this section, we develop a reliability-aware placement based  genetic algorithm that jointly optimizes node mapping and routing while Considering the reliability of the SFCs to achieve desirable reliability with minimum resource consumption. The pseudo code of the proposed genetic algorithm is provided in Algorithm~\ref{alg:genetic}. The algorithm finds solutions in the processes of initial population generation, fitness evaluation, selection, crossover, and mutation. First, generate random population of P individuals and evaluate the fitness of each individual in the population and select set of parent individual from the population according to their fitness (lines 1 through 5 of Algorithm~\ref{alg:genetic}). According to a crossover probability, cross over the parents to form new offspring and with a mutation probability mutate new offspring then checking for the new individuals satisfy the constraints and update people ranks (lines 6 through 21 of Algorithm~\ref{alg:genetic}). If it converges converge provide fittest individual and terminate possess else this possess repeats as far as converges converge (lines 22 through 27 of Algorithm~\ref{alg:genetic}). In the following sub-sections we present the encoding mechanism, feasibility checking process, and the fitness function.
    
\subsection{Encoding mechanism}\label{subsec:encod}
    In general, our NFV Network encodes two chromosomes: chromosome~1 (Table~\ref{tab:chr1VNF} and~\ref{tab:chr1SFC}) represents the location and assigned backups of VNFs of each service also used link for connecting VNFs of each service and chromosome~2 (Table~\ref{tab:chr2}) represents the location, function type, list of user VNFs and calculated reliability of each backup VNF in the network. Given a network with i services, j VNFs and k Backup VNFs, chromosome~1 consists of i+j genes, i genes for representing the path of each service and i gene for representing the properties and assigned backups of a primary VNF. Chromosome~2 consists of k genes, each of which represents the specifications of a Backup VNF.
    \begin{table*}
    	\caption{Chromosome 1: VNF’s genes.}\label{tab:chr1VNF}
        \centering
        \begin{tabular}{|p{0.08\linewidth}|p{0.09\linewidth}|p{0.06\linewidth}|p{0.1\linewidth}|p{0.22\linewidth}|p{0.16\linewidth}|p{0.09\linewidth}|}	
    	\hline
    	    {Location} & {Function type} & {SFC id} & {Position in SFC} & {Used links to access assigned backup VNFs} & {List of assigned backup} & {Reliability} \\\hline
    	\end{tabular}
    \end{table*}
    \begin{table*}
    	\caption{Chromosome 1: SFC's genes.}\label{tab:chr1SFC}
        \centering
        \begin{tabular}{|p{0.08\linewidth}|p{0.09\linewidth}|p{0.28\linewidth}|p{0.44\linewidth}|}	
    	\hline
    	    {SFC id} & {Used links} & {Maximum tolerable delay} & {Minimum tolerable reliability} \\\hline
    	\end{tabular}
    \end{table*}
    \begin{table*}
    	\caption{Chromosome 2.}\label{tab:chr2}
        \centering
        \begin{tabular}{|p{0.08\linewidth}|p{0.12\linewidth}|p{0.25\linewidth}|p{0.44\linewidth}|}	
    	\hline
    	    {Location} & {Function type} & {List of user VNF} & {Reliability} \\\hline
    	\end{tabular}
    \end{table*}
\subsection{Checking for feasibility}\label{subsec:feas}
    Crossover phase and mutilation phase can cause VNF mapping that cannot satisfy present constraints in previous section. So the phase added after crossover and mutilation to check the feasibility of mapping results. If there are invalid mapping result, we have to correct them so that they can meet all the constraints.
\subsection{Selection}\label{subsec:selec}
    In selection strategy, we use a ranking scheme to avoid premature convergence. The ranking scheme is such that when the fitness value of each individual calculated, the individuals are sorted based on this value then the individuals based its rank selected for crossover phase, which means the individual with larger fitness value has a higher chance of taking part in crossover presses.
    
    The penalty process is used to calculate points and rank. In this process the individuals who do not satisfy the constraints of the problem, are given a negative score depending on the importance of the violated constraint. This makes the individual violating constraints get lower scores, thus reducing the probability of selecting the individual in the next crossover.
\subsection{Convergence Condition}\label{subsec:conv}
    To evaluate the performance of the algorithm we modify the degree of diversity~\cite{gao2015virtual} as follow:
    \begin{align}
        D_p = \frac{2}{(P(P-1))} \sum_{p_1=1}^{P-1}{\sum_{p_2 = p_1 + 1}^P}{\frac{\abs{F_{p_1} - F_{p_2}}}{F_{max}}}\label{eqR25}
    \end{align}
    where $\abs{F_{p_1}-F_{p_2}}$ is the absolute difference of the fitness of individual $p_1$ and $p_2$; and $F_{max}$ is the maximum fitness value in the generation. For 5 generations or more, if $D_p$ value is less than a given, we consider that the algorithm is converged.
    
    \begin{algorithm}
    	\caption{Pseudo-Code of Reliability-aware and minimum-Cost based Genetic (RCG)}
    	\label{alg:genetic}
    	\allowdisplaybreaks
    	\begin{algorithmic}[1]
        	\INPUT{ $\alpha$, $\beta$, $\gamma$, $\delta$}
        	\Statex{\texttt{$\alpha$: size of population}}
        	\Statex{\texttt{$\beta$: rate of elitism}} 
        	\Statex{\texttt{$\gamma$: rate of mutation}}
        	\Statex{\texttt{$\delta$: convergence threshold}}
        	\OUTPUT{X}
        	\Statex{\texttt{X: solution}}
        	\Statex{\texttt{//Initialization}}
        	\State{generate $\alpha$ feasible solutions randomly;}
        	\State{save them in the population Pop;}
        	\Statex{\texttt{//Loop until the Convergence Condition}}
        	\For{$cent=0; cent\leq threshold; cent$\textit{++}}
        	    \Statex{\texttt{//Elitism based selection}}
        	    \State{number of elites $ne = \alpha\cdot \beta$;}
        	    \State{select the best $ne$ solutions in Pop and save them in $Pop_1$;}
        	    \Statex{\texttt{//Crossover}}
        	    \State{number of crossover $nc = (\alpha - ne)/2$;}
        	    \For{$j=1$ to $nc$}
        	        \State{select two solutions $X_A$ and $X_B$ from $Pop$;}
        	        \State{generate $X_C$ and $X_D$ by one-point crossover to $X_A$ and $X_B$;}
        	        \State{save $X_C$ and $X_D$ to $Pop_2$;}
        	    \EndFor
        	    \Statex{\texttt{//Mutation}}
        	    \For{$j=1$ to $nc$}
        	        \State{select a solutions $X_j$ from $Pop_2$;}
        	        \State{mutate each bit of $X_j$ under the rate $\gamma$ and generate a new solution $X_j^\prime$;}
        	        \If{$X_j^\prime$ is unfeasible}
        	            \State{update $X_j^\prime$ with a feasible solution by repairing $X_j^\prime$}
        	       \EndIf
        	       \State{update $X_j$ with $X_j^\prime$ in $Pop_2$;}
        	    \EndFor
        	    \Statex{\texttt{//Updating}}
        	    \State{calculate rank of $Pop_2$ individuals;}
        	    \State{update $Pop=Pop_1+Pop_2$;}
        	    \Statex{\texttt{//Convergence Condition}}
        	    \State{calculate $D_p$ using Eq.~\ref{eqR25};}
        	    \If{$D_p$ value is less than $\delta$}
        	        \State{break;}
        	   \EndIf
        	\EndFor\\
        	\Return{best solution $X$ in $Pop$;}
    	\end{algorithmic}
	\end{algorithm}
    
\section{Performance Evaluation}\label{sec:results}
\begin{figure}[!htbp]
    \centering
        \begin{subfigure}{0.49\textwidth}
    	    \centering
    		\includegraphics[width=0.8\linewidth]{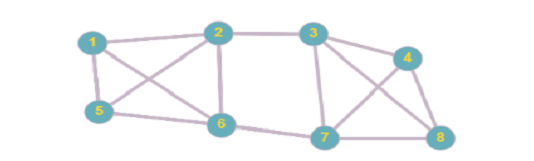}
        	\caption{Sample network with 8-nodes (refereed as 8-node network).}
        	\label{fig:20nodetopo}
    	\end{subfigure}
    	\begin{subfigure}{0.49\textwidth}
    	    \centering
    		\includegraphics[width=0.8\linewidth]{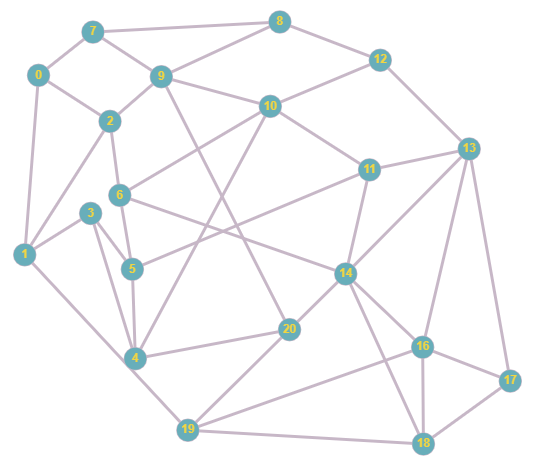}
        	\caption{Sample network with 20-nodes (refereed as 20-node network).}
        	\label{fig:8nodetopo}
    	\end{subfigure}
    	\caption{Network Topologies.}
        \label{fig:nettopo}
    \end{figure}

    In this section, we evaluate the performance of the proposed MILP model and the heuristic algorithms. IBM CPLEX is used to solve the mathematical formulation of problem and Python to implement the heuristic algorithms. All of the simulations are done using a machine with 2.30 GHz Intel Xeon CPU and 16 GB RAM. Two substrate networks are considered:
    \begin{itemize}
        \item An 8-node and 14-link NSF network and hosting 4 network services (Fig.~\ref{fig:8nodetopo}). 
        \item A 20-node and 40-link NSF network and hosting 5 to 30 network services. (Fig.~\ref{fig:20nodetopo}). 
    \end{itemize}
    We assume there are four types of function in the network, and each physical node can execute up to 4 VNFs. The reliability of nodes are specified randomly using uniform distribution between 0.90 and 0.96. It is assumed that each SFC requires 3 VNFs and a minimum reliability of 0.98. This requirement is not hold in random placement. The link delay and bandwidth are respectively fixed to 10ms and 20~(units).
    We consider four different demand scenarios to evaluate the proposed solution:
    
    \begin{enumerate}
        \item $f_3\rightarrow f_2\rightarrow f_1 (\sigma_1=8,\delta_1=5,b_1=2,\phi_1=50)$
        \item $f_2\rightarrow f_1\rightarrow f_3 (\sigma_2=1,\delta_2=3,b_2=4,\phi_2=50)$
	    \item $f_2\rightarrow f_3\rightarrow f_1 (\sigma_3=1,\delta_3=8,b_3=2,\phi_3=60)$
	    \item $f_1\rightarrow f_3\rightarrow f_2 (\sigma_4=3,\delta_4=5,b_4=4,\phi_4=60)$
    \end{enumerate}
    For the purpose of performance comparison and bench-marking, three additional schemes are implemented. They are:
    \begin{enumerate}
        \item A dedicated protection (DP): as for DP, we exploit state-of-the-art solution proposed in \cite{qu2017reliability} for resource allocation problem. The mentioned solution exploits dedicated protection to guarantee the end-to-end reliability of services.
        \item A none protection (NP):this algorithm is reliability unaware, i.e., no any backup VNF should be defined through this solution. We removed the reliability constraints from our solution and use it as NP. 
	    \item A Random placement (RP):this random placement algorithm neither satisfies the reliability requirements nor the end-to-end delay constraints. This makes the algorithm to have the optimal response time when deploying SFCs. RP algorithm only focuses on the routing constraints while randomly doing (both primary and backup) VNFs placement.
    \end{enumerate}
    In the first step, all of the above-mentioned algorithms are tested on 8-nodes network and the results are shown in Table~\ref{tab:routeRes}. According to the table, NP achieves the smallest bandwidth utilization $10.0\%$ whle RP has the smallest execution time. This happens because the main objective of NP is to minimize the bandwidth consumption without considering any backup VNFs. Similarly, the main objective of RP is to simplify the management process. It should be mentioned that these two algorithms fail to satisfy the reliability constraints. On the other hand, SP-MILP, RCG, and DP satisfy the reliability constraints of the network services at the cost of increasing the bandwidth utilization. Due to this reason, SP-MILP achieves minimum bandwidths utilization $9.28\%$ along with maximum execution time of $104$ and $238$s, respectively. Conversely, reliability of DP is the highest among all. The bandwidth utilization achieved by RCG are better than DP solution. As it is expected, the execution time of RCG is significantly lower than SP-MILP. In the next step, we test the algorithms on a 20-nodes network which is hosting 5 to 30 network services. We compare the performance of the algorithms through five metrics: I. Reliability, II. Execution time (CPU time), III. Computational resource consumption (CPU utilization), IV. Link utilization, and V. Computational complexity (order of complexity).
    
    \begin{table*}[t]
    	\caption{Routing Results (8-Nodes Network).}\label{tab:routeRes}
        \centering
        \scriptsize
        \begin{tabular}{|p{0.06\linewidth}|p{0.02\linewidth}|p{0.45\linewidth}|p{0.07\linewidth}|p{0.08\linewidth}|p{0.06\linewidth}|}	
    	\hline
    	    \textbf{Algorithm} & \textbf{NS} & \textbf{Routing and VNFs assignment (VM (VNFs))} & \textbf{Reliability} & \textbf{Bandwidth Utilization} &  \textbf{CPU time (s)}\\\hline\hline
    	    \multirow{4}{*}{MILP} & 1 & $5\rightarrow\{6,2,1\}(f_3 )\rightarrow\{5,4,7\}(f_2 )\rightarrow\{8,7,2\}(f_1)$ & 0.994 & \multirow{4}{*}{49.28$\%$} & \multirow{4}{*}{238.922} \\\cline{2-4}
    	    & 2 & $1\rightarrow\{8,4,7\}(f_2)\rightarrow\{3,7,2\}(f_1 )\rightarrow\{5,2,1\}(f_3)\rightarrow 3$ &   0.994 & &  \\\cline{2-4}
    	    & 3 & $1\rightarrow\{3,4,7\}(f_2 )\rightarrow\{8,2,1\}(f_3 )\rightarrow\{6,7,2\}(f_1 )\rightarrow 8$ & 0.995 & &  \\\cline{2-4}
    	    & 4 & $3\rightarrow\{5,7,2\}(f_1 )\rightarrow\{3,2,1\}(f_3 )\rightarrow\{6,4,7\}(f_2 )\rightarrow 5$ & 0.994 & &  \\\hline\hline
    	    \multirow{4}{*}{Genetic} & 1 & $8\rightarrow 7\rightarrow\{6,1,5\}(f_1 ),\{6,3\}(f_2 )\rightarrow 7,\{4,6\}(f_3)\rightarrow 7\rightarrow 6\rightarrow 5$ & 0.998 & \multirow{4}{*}{66.71$\%$} &  \multirow{4}{*}{0.639} \\\cline{2-4}
    	    & 2 & $1\rightarrow\{5,3,1\}(f_1 )\rightarrow 2\rightarrow\{3,6\}(f_2 )\rightarrow\{7,5\}(f_3)\rightarrow 3$ & 0.998 & &  \\\cline{2-4}
    	    & 3 & ${1,3}(f_2 )\rightarrow\{5,3\}(f_3 ),\{5,6,3\}(f_1)\rightarrow 6\rightarrow 7\rightarrow 8$ & 0.999 & & \\\cline{2-4}
    	    & 4 & $3\rightarrow\{8\}(f_1 )\rightarrow 3\rightarrow 2\rightarrow\{1\}(f_2 ),\{1\}(f_3)\rightarrow 5$ & 0.999 & & \\\hline\hline
    	    \multirow{4}{0.06em}{NO protection} & 1 & $8\rightarrow 3\rightarrow 2(f_3)\rightarrow 1(f_2 )\rightarrow 5(f_1 )$ & 0.796 & \multirow{4}{*}{10.00$\%$} &  \multirow{4}{*}{0.031} \\\cline{2-4}
    	    & 2 & $1(f_2) \rightarrow 5(f_1) \rightarrow 2(f_3)\rightarrow 3$ & 0.810 & &  \\\cline{2-4}
    	    & 3 & $1(f_2) \rightarrow 2(f_3) \rightarrow 5(f_1) \rightarrow 6 $ & 0.797 & & \\\cline{2-4}
    	    & 4 & $3\rightarrow 2(f_1)\rightarrow 1(f_3)\rightarrow 5(f_2)$ & 0.795 & & \\\hline\hline
    	    \multirow{4}{*}{Random} & 1 & $8\rightarrow 3\rightarrow 2\rightarrow 5(f_1),5(f_2)\rightarrow 2\rightarrow 3\rightarrow 8(f_3)\rightarrow 3\rightarrow 2\rightarrow 5$ & 0.809 & \multirow{4}{*}{63.60$\%$} &  \multirow{4}{*}{0.023} \\\cline{2-4}
    	    & 2 & $1\rightarrow 6\rightarrow 7\rightarrow\{4,7\}(f_1 ),\{4,8\}(f_2 )\rightarrow\{3,2\}(f_3)$ & 0.998 & &  \\\cline{2-4}
    	    & 3 & $1\rightarrow\{5,3\}(f_1 )  5\rightarrow 1(f_2)\rightarrow\{2,3,8\}(f_3)\rightarrow 3\rightarrow 8$ & 0.908 & & \\\cline{2-4}
    	    & 4 & $3\rightarrow 7\rightarrow\{6,3\}(f_1)\rightarrow 5(f_2),5(f_3)$ & 0.868 & &  \\\hline\hline
    	    \multirow{4}{0.06em}{Dedicate protection} & 1 & $8 \rightarrow\{3,2\}(3)\rightarrow 2 \rightarrow\{7,4\}(2)\rightarrow\{5,6\}(1) $ & 0.982 & \multirow{4}{*}{62.85$\%$} &  \multirow{4}{*}{0.023} \\\cline{2-4}
    	    & 2 & $1\rightarrow\{7,4\}(2)\rightarrow\{3,6\}(1)\rightarrow\{1,2\}(3) \rightarrow 2 \rightarrow 3$ & 0.986 & &  \\\cline{2-4}
    	    & 3 & $1 \rightarrow\{3,8\}(2)\rightarrow\{7,2\}(3)\rightarrow 2 \rightarrow\{1,6\}(1)\rightarrow 8$ & 0.986 & &  \\\cline{2-4}
    	    & 4 & $3\rightarrow\{5,6\}(1)\rightarrow\{7,2\}(3)\rightarrow\{3,4\}(2)\rightarrow 5$ & 0.982 & &  \\\hline
    	\end{tabular}
    \end{table*}
    
\subsection{Reliability}
\begin{figure}[!htbp]
        \centering
    	\includegraphics[width=\linewidth]{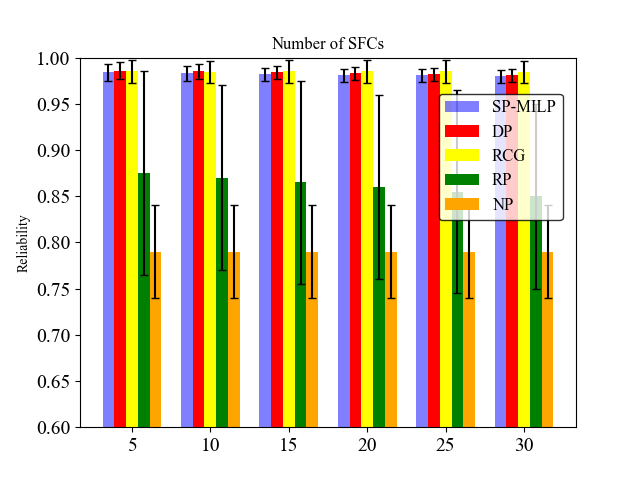}
    	\caption{Reliability versus the number of requested services}
        \label{fig:6}
    \end{figure}

    Reliability is the ability of the network (including routing and processing devices) to consistently perform its intended or required function, on demand and without degradation or failure. It is critical in many case to reduce the probability of  failure occurrence that could cause the entire service presence to come crashing down. In this part, we compare the reliability of the proposed resource allocation algorithms with state-of-the-art algorithms.
    The first group of carried out tests aims to evaluate and compare the achieved reliability of above-mentioned algorithms. In our emulation, we consider the minimum acceptable reliability to be 0.98 for each SFCs, however, RP and NP cannot manage to achieve this reliability. Based on simulation results, the achieved reliability of proposed algorithms are reported in Fig.~\ref{fig:6}. 
    In each scenario, we increase the number of available SFCs in the network from 5 SFCs up to 30 SFCs by adding 5 new service function chains in each iteration and calculating the average SFCs reliability. As can be seen, achieved reliability for NP and RP in all scenarios are lower than SP-MILP, RCG, and DP. This happens because these algorithms do not consider the service reliability as metric so they cannot meet the reliability constraints. Considering the error bar in Fig.~\ref{fig:6}, NP and RP algorithms have a very high range of results in term of reliability. In contrast, the other algorithms (DP, SP-MILP, and RCG) not only satisfy the reliability constraint but also they achieve a system reliability higher than 0.98 in most cases. It should be mentioned that DP, RCG, and SP-MILP try to find the minimum reliability higher than acceptable reliability to reduce the total waste of resources. Therefore, the lower error band (distance from desirable reliability) is more intended. Based on our simulation results, SP-MILP has the most stable outcomes around the desirable reliability. This is due to the fact that SP-MILP finds the optimal solution for a system with a reliability higher than a pre-defined threshold but with lowest computational resource consumption.

\subsection{Execution time (CPU time)}
\begin{figure*}[!htbp]
    \centering
        \begin{subfigure}{0.49\textwidth}
    	    \centering
    		\includegraphics[width=1\linewidth]{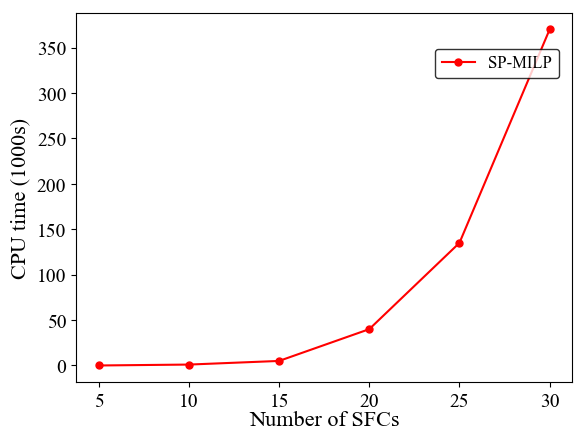}
        	\caption{Optimal solution.}
        	\label{fig:2a}
    	\end{subfigure}
    	\begin{subfigure}{0.49\textwidth}
    	    \centering
    		\includegraphics[width=1\linewidth]{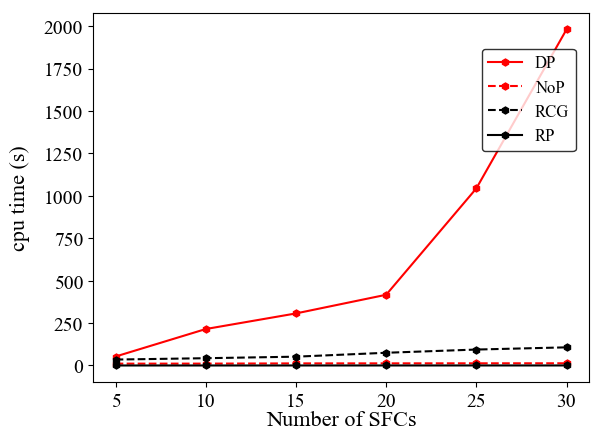}
        	\caption{Heuristic algorithms.}
        	\label{fig:2b}
    	\end{subfigure}
    	\caption{CPU time versus number of network services (20-nodes network).}
        \label{fig:2}
    \end{figure*}

    In a general sense, high-performance algorithm means getting the most out of the resources. This translates to utilizing the CPU as much as possible. Consequently, CPU utilization becomes a very important metric to determine how well an algorithm is using the computational resources. Talking about a predefined goal, high-performance algorithm uses less resource to achieve the goal in compare to less productive ones. In this way, CPU time (or execution time) is defined as the time spent by the system executing each algorithm, including the time spent executing run-time or system services on its behalf. In Fig.~\ref{fig:2}, we evaluate the execution time of preferred algorithms for different service chain request, i.e. execution time versus the number of requested services (VNFs) is depicted. It is worth mentioning that RP and NP are not reliability-aware while RCG, DP, and SP-MILP are. Since the execution time of SP-MILP is dramatically higher than other methods, we put it in a separate plot to keep the plots clear and simple to read. In this way, the execution time of SP-MILP over 20-nodes network is reported in Fig.~\ref{fig:2a} while the other algorithms are measured in Fig.~\ref{fig:2b}. According to Fig.~\ref{fig:2a}, SP-MILP is too complex for even small-scale networks, therefore, it is not applicable for real-world scenarios. However, the reliability of the solution provided by SP-MILP is the optimal one. On the other hand, RP and NP methods have a very low cpu time which makes them applicable for real-world networks. However, both methods do not satisfy some of the constraints of the problem, and this allows them to respond quickly as compared to other methods.
    Considering reliability-aware solutions, DP has a medium-high execution time meaning dramatically lower than SP-MILP but sufficiently higher than NP, RP, and RCG. Consequently, we can conclude that although DP considers the reliability in allocation of resources, due to its high execution time it is not practical for medium and large scale networks. Comparing the proposed genetic algorithm with DP, NP, RP, and MILP, not only RCG is reliability aware but also it is practical and could be used for real-world scenarios.
    
\subsection{Computational resource consumption (CPU utilization)}
\begin{figure}[!htbp]
        \centering
    	\includegraphics[width=\linewidth]{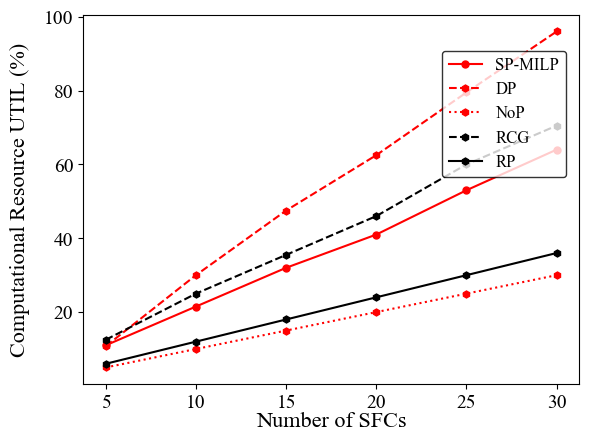}
    	\caption{CPU utilization ($\%$) versus the number of requested services}
        \label{fig:3}
    \end{figure}

    Although CPU time (execution time) is a very important metric to measure the performance of algorithms, it is not a comprehensive metric. As an example, consider an algorithm that exploits 20$\%$ of CPU capacity after a period of 100ms execution time and another algorithm which exploits 90$\%$ of CPU after a period of 90ms computation. This means that CPU time should be measured along with computational resource consumption. Computational resource consumption (or CPU utilization) is the sum of work handled by CPU and is used to estimate system performance. As mentioned in the above example, performance of the algorithm can vary according to both the amount of computing and the execution time of the algorithm. In this sub-section, we evaluate the computational resource consumption of each preferred method. This includes the total resources required for the primary functions and the backup functions for each of the mentioned resource allocation algorithm. To this end, Fig.~\ref{fig:3}, shows the CPU utilization versus the number of requested services (VNFs). 
    Similar to previous sub-section, RP and NP have the lowest CPU utilization, however, they are reliability unaware. This is mean that they reduce consumption of the computational resource at the cost of an intense reduction in the system reliability. Among reliability aware solutions, SP-MILP has the lowest CPU utilization with the sacrifice of CPU time. This means that although the CPU utilization is lower than DP and RCG, it is not practical due to its high execution time. Comparing RCG and SP-MILP, it is clear that the meta-heuristic method closely follows the optimal solution obtained via the mathematical optimization model. Comparing DP and RCG, dedicated protection scheme requires more computational resources than shared resource consumption. It also needs a higher CPU time which clearly explains why RCG is superior to DP in terms of both CPU time and utilization.
    
\subsection{Link utilization results}
\begin{figure}[!htbp]
        \centering
    	\includegraphics[width=\linewidth]{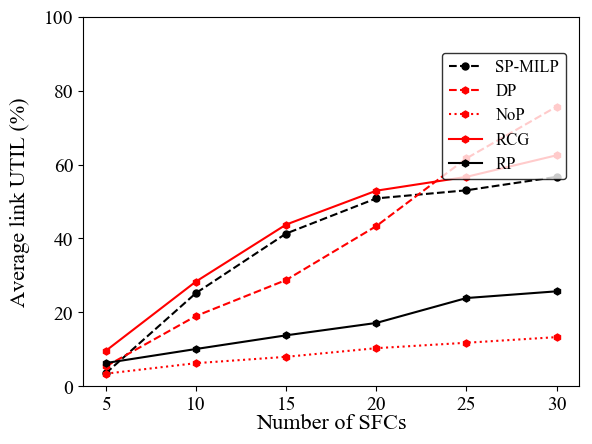}
    	\caption{Bandwidth utilization ($\%$) versus number of network services (20-nodes network).}
        \label{fig:4}
    \end{figure}

    Bandwidth utilization is one of the most basic and critical statistics available in assessing a network resource allocator. It shows the average traffic levels on links compared to total capacity of those lins.
    Fig.~\ref{fig:4} shows the comparison of the average link utilization between the RCG and other algorithms. We can comprehend how RP introduces a lot of overload links. According to this plot, maximum link utilization of NP method is much lower than other methods.
    Sharing protection methods involving the SP-MILP and the genetic algorithm initially consume more bandwidth than DP method, but with increasing number of SFC, bandwidth consumption of the DP method will increase in sharing protection methods. Another issue that can be inferred from Fig.~\ref{fig:4} is that bandwidth consumption in the RCG is very close to the SP-MILP.
    Initially, due to the small number of primary functions, the intensity of sharing the backup functions is low as result, dedicated methods less bandwidth consuming, But with the increase in the number of primary functions and the upgrading of the intensity of sharing the backup functions and decrease the need for more backup functions, the required bandwidth of  shared methods reduced.
    
\subsection{Computational Complexity (order of complexity)}
    DP and SP-MILP implemented as a mixed-integer linear programming model. It is stated in ~\cite{till2003empirical} that the complexity of MILP algorithms grows with increasing problem size and presents a table that contains the relationship between the number of variables and the computational complexity.
    According to the table, since the DP variables are 100 to 10,000, its computational complexity is $O(n^2)$ and also the SP-MILP variables are more than 10,000, its computational complexity is $O(n^4)$.

\section{Conclusion}\label{sec:conclusions}
    In order to provide reliable service function chains, a large number of backup functions are required. Although this redundancy is essential, it may sufficiently reduce the network resource efficiency if resources are not well assigned. To solve this problem, we exploited a Shared Protection (SP) algorithm with Active-Standby redundancy as an optimization issue to achieve the optimal network in the virtualization environment. We first formulated the problem as a mixed-integer linear programming (MILP) and found the optimal solution of the problem then we compared the SP-MILP method with three other methods: Dedicated Protection (DP), No Protection (NP) and Random Placement (RP). SP-MILP has a very high time complexity compared to the other approaches. But in terms of Computational resource consumption, it is about $33\%$ lower than DP. Also, bandwidth consumption in the case of a high number of services is $25.2\%$ lower than DP. To solve the complexity problem, we proposed a genetic algorithm to solve the aforementioned problem. Based on simulation results, the proposed genetic algorithm with time complexity yields an optimize gap of approximately $9\%$ bandwidth consumption and $9.3\%$ optimize gap in computational resource consumption, compared to the SP-MILP response. 
        
\bibliographystyle{elsarticle-num}

\bibliography{references}

\clearpage

\section*{Biographies}\label{sec:12}

\noindent\begin{minipage}{0.2\textwidth}
\includegraphics[width=1.15in,height=1.15in,clip,keepaspectratio]{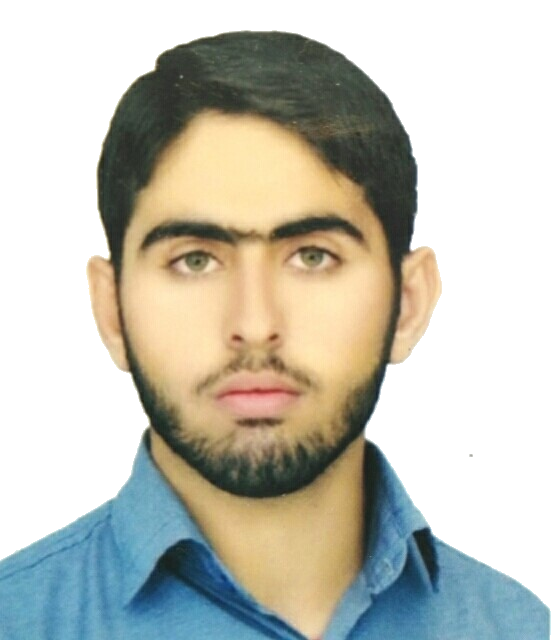}
\end{minipage}%
\hfill%
\begin{minipage}{0.82\textwidth}
\textbf{Abolfazl Ghazizadeh} received the bachelor’s degree in Computer engineering (Hardware) from Hakim Sabzevari University. He earned a master’s degree at Tarbiat Modares University with a focus on optimization in Network function virtualization. His previous research mainly dealt with Computer Networks, Network QOS,Network optimization ,service function chaining, software-defined networking (SDN)
\end{minipage}%

\hfill \break

\noindent\begin{minipage}{0.2\textwidth}
\includegraphics[width=1.15in,height=1.15in,clip,keepaspectratio]{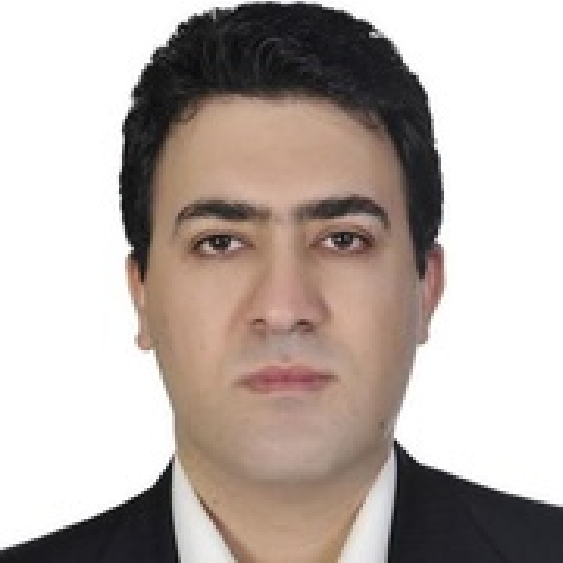} 
\end{minipage}%
\hfill%
\begin{minipage}{0.82\textwidth}
\textbf{Behzad Akbari} received the B.S., M.S., and PhD degree in computer engineering from the Sharif University of Technology, Tehran, Iran, in 1999, 2002, and 2008 respectively. His research interest includes Computer Networks, Multimedia Networking Overlay and Peer-to-Peer Networking, Peer-to-Peer Video Streaming, Network QOS, Network Performance Analysis, Network Security, Network Security Events Analysis and Correlation, Network Management, Cloud Computing and Networking, Software Defined Networks.
\end{minipage}%

\hfill \break

\noindent\begin{minipage}{0.2\textwidth}
\includegraphics[width=1.15in,height=1.15in,clip,keepaspectratio]{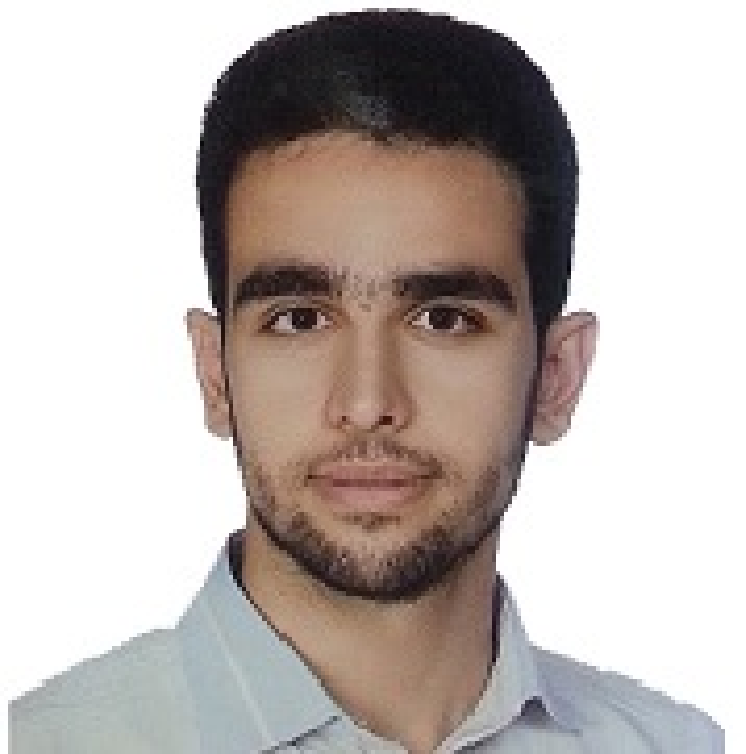}
\end{minipage}%
\hfill%
\begin{minipage}{0.82\textwidth}
\textbf{Mohammad M. Tajiki}
is a research associate at Queen Mary University of London, UK, holding two PhD degrees one from University of Rome Tor Vergata in Electrical Engineering and another one from Tarbiat Modares University in Computer Engineering. His main research interests are Network Monitoring, Network Function Virtualization, Network QoS, data centre networking, traffic engineering, service function chaining, IPv6 segment routing, and software-defined networking (SDN).
\end{minipage}%

\hfill \break

\end{document}